\documentclass[aps,pra,preprint]{revtex4}      

\def\bd{\begin{displaymath}}
\def\be{\begin{equation}}
\def\ed{\end{displaymath}}
\def\ee{\end{equation}}

\newcommand{\sn}{{\rm sn}}
\newcommand{\dn}{{\rm dn}}

\usepackage[dvips]{graphicx}
\usepackage{psfrag}

\begin{document}

{
\title{\bf Exact nonstationary solutions to the 
mean-field equations of motion for two-component Bose-Einstein condensates in periodic potentials} 
\author{R. Mark Bradley}
\affiliation{
Department of Physics,
Colorado State University,
Fort Collins, CO 80523 USA
}
\author{Bernard Deconinck and J. Nathan Kutz}
\affiliation{
Department of Applied Mathematics,
University of Washington,
Seattle, WA 98195 USA
}

\begin{abstract}

We study the dynamics of two-component Bose-Einstein condensates in periodic
potentials in one dimension.  Elliptic potentials which
have the sinusoidal optical potential as a special case are considered.  We construct exact 
nonstationary solutions to the mean-field equations of motion. 
Among the solutions are two types of temporally-periodic solutions ---
in one type there are condensate oscillations between neighboring potential wells, while in the other the condensates oscillate from side to side within the wells.
Our numerical studies of the stability of these
solutions suggests the existence of one-parameter families of stable 
nonstationary solutions.

\end{abstract}
\maketitle
}

{\bf I. Introduction}

In conventional magnetic traps, the spins of the alkali atoms making up a
Bose-Einstein condensate (BEC) are frozen, and all atoms  are in the first
hyperfine manifold.  Thus, even if the atoms have spin, the condensate order
parameter $\psi$ is a single complex scalar.  

Over the last few years, two methods to produce mixtures of two distinguishable
BEC's have been developed and implemented.  In these condensate mixtures,
$\psi$ has two components. The Boulder group, for example, produced a mixture
of two condensates consisting of two different hyperfine spin states of
$^{87}$Rb \cite{Myatt,Hall}.  Initially, a single condensate in the spin state
$\vert 1 \negthinspace>\>\negthinspace$ with hyperfine spin $F=1$ was trapped
magnetically.  A two-photon transition was then used to transfer a portion of
the atoms to a second spin state $\vert 2 \negthinspace>\>\negthinspace$ with
$F=2$.  Once the two-photon pulse had ended, the number of atoms in each
condensate was essentially constant. Modugno {\sl et al.} \cite{Modugno}, on
the other hand,  produced a mixture of two Bose-Einstein condensates of
different atomic species, $^{41}$K and $^{87}$Rb \cite{spinor}.

An atom placed in a standing light wave is subject to induced dipole forces. 
The resulting atomic potential is sinusoidal, and is referred to as an optical
lattice.   Merely by altering the phase, wavelength and intensity of the light,
the position, lattice spacing and well depth of the optical lattice can be
adjusted.   Thus, optical lattices are well-characterized and controllable.   

By adiabatically increasing the depth of the lattice potential, trapped BEC's
can be transferred into an optical lattice.  BEC's on optical lattices have
attracted a great deal of interest for several reasons. In experiments, the
analogs of the ac and dc Josephson effects \cite{Josephson},  Bloch
oscillations and Landau-Zener tunneling \cite{Morsch}, number-squeezed states
\cite{Orzel},  and the Mott-insulator transition \cite{Greiner} have been
studied.  In addition, an optical lattice in which each site is occupied by a
single alkali atom in its ground state is a promising candidate for a register
in a quantum computer.  One route currently being explored to initialize such a
register is to begin with a BEC and then adiabatically turn on an optical
lattice, with the result that one atom in its  ground state occupies each site
\cite{Porto}.  From a theoretical standpoint, BEC's on optical lattices are
interesting because the simple sinusoidal form of the potential has made it
possible to obtain some exact solutions to the mean-field equations of motion,
in spite of their nonlinearity
\cite{Bronski_1,Bronski_2,Deconinck_1,Deconinck_2}.

Two condensates confined to a cigar-shaped trap are quasi-one-dimensional
if the trap is highly elongated and if the tranverse dimensions of the trap are
comparable to the healing lengths.  A quasi-one-dimensional optical lattice  can
be realized by superimposing a standing light wave on the cigar-shaped trap.   In
the mean-field approximation, the condensates are described  by coupled
nonlinear Schr\"odinger equations in one space dimension with periodic external
potentials.

In this paper, we study the dynamics of two-component condensates in periodic
potentials in one dimension (1D).  We will consider elliptic potentials which
have the sinusoidal potential as a special case.  Exact stationary solutions to
this problem have been found  by Deconinck {\sl et al.}
\cite{Deconinck_1}.  We will use these solutions to construct exact {\it
nonstationary} solutions to the problem.  Our numerical studies of the stability of these
solutions strongly suggest that there are one-parameter families of stable
solutions.

We will only study condensate mixtures in which the atoms of the different
components have the same atomic mass $\mu$. This means that mixtures of
condensates of different atomic species are excluded from our analysis, and
accordingly we will confine ourselves to consideration of the two-component
condensates like those studied by the Boulder group \cite{Myatt,Hall}.  

The paper is organized as follows.  In Section II, we introduce the mean-field
equations of motion and develop a general method of constructing nonstationary
solutions.  This method is applied to condensates in a 1D elliptic potential in
Sec. III, and the physical interpretation of the resulting solutions is
discussed in Sec. IV. The stability of the solutions is the subject of Sec. V. 
Finally, we summarize our results in Sec. VI.

{\bf II. General Development}

In the two-component condensates studied by the Boulder group, the number of
atoms in each condensate is  nearly constant \cite{Myatt,Hall}.  We will therefore neglect loss
of atoms to the normal phases and switching of atoms from one condensate to
another.  The mean-field equations of motion are then
\be 
i\hbar{{\partial\psi_j}\over {\partial t}} = 
-{{\hbar^2}\over {2 \mu}}{{\partial^2 \psi_j}\over{\partial x^2}}
+ \left (\sum_{l=1}^2 \alpha_{jl}\vert\psi_l\vert^2 \right )\psi_j
+ V_j\psi_j
\label{GP}
\ee 
for $j=1, 2$.  Here $\psi_j = \psi_j(x, t)$ is the condensate wave
function for the $j$th species and $\alpha_{jl}$ describes the interaction of
an atom in the $j$th condensate and  an atom in the $l$th condensate. The
$2\times 2$ symmetric matrix $M=\{\alpha_{ij}\}$ will be referred to as the
interaction matrix.  The interaction strengths $\alpha_{11}$, $\alpha_{12}$
and $\alpha_{22}$ are known to the $1\%$ level for $^{87}$Rb, and are in the
proportion 1.02 : 1 : 0.97 \cite{Burke}.  We shall neglect the difference between the
interaction strengths and set $\alpha = \alpha_{11} = \alpha_{12} =
\alpha_{22}$.

In the case of an optical lattice, the potentials $V_1$ and $V_2$ are
sinusoidal.   
$V_1$ and $V_2$ are very nearly equal for linearly polarized light, provided
that the detuning is not too small \cite{Roberts}.  We shall neglect the difference $V_1-V_2$ and set 
\be
V\equiv V_1 = V_2 = -V_0\sin^2 qx ,
\label{optical_potential}
\ee
where $V_0$ is a constant which depends on the intensity of the light, $q\equiv 2\pi/\lambda$ and $\lambda$ is the optical wavelength.  In
fact, we will study the more general potential
\be
V = -V_0 \,{\rm sn}^2(qx, k),
\label{elliptic_potential}
\ee
where ${\rm sn}\,(qx, k)$ is the Jacobian elliptic sine function with elliptic
modulus $k\in [0,1]$. The potential (\ref{elliptic_potential}) reduces to the
optical potential (\ref{optical_potential}) for $k=0$ and to the single
potential well (or barrier) $V = -V_0 \tanh^2 qx$ for $k=1$.  For $0< k < 1$,
the potential is periodic with period $2K(k)/q$, where
\be
K(k)\equiv \int_0^{\pi/2}{{d\theta}\over{\sqrt{1-k^2\sin^2\theta}}}.
\ee
Plots of the potential (\ref{elliptic_potential}) for a range of $k$ values 
may be found in Ref. \cite{Bronski_1}. To simplify the notation, we will not
display the $k$ dependence of the Jacobian elliptic functions from this point
on.

Let ${\vec\psi}\equiv (\psi_1, \psi_2)^T$.  With the simplifications we have made, the equation of motion is 
\be
i\hbar{{\partial{\vec\psi}}\over {\partial t}} = -{{\hbar^2}\over {2 \mu}} {{\partial^2 \vec \psi}\over{\partial x^2}} 
+\alpha ({\vec\psi}^\dagger{\vec\psi}){\vec\psi} - V_0 {\rm sn}^2 (qx)\, {\vec\psi}.
\label{original_eom}
\ee
This equation is further simplified by introducing the dimensionless variables
$\tilde x = qx$, $\tilde t = \hbar q^2 t/\mu$, $\tilde\psi_j = (\sqrt{\vert\alpha\vert\mu}/\hbar q)\psi_j$,
$\tilde\alpha = {\rm sgn}(\alpha)$ and $\tilde V_0 = \mu V_0/(\hbar q)^2$ and then dropping the tildes.
This gives 
\be
i{{\partial{\vec\psi}}\over {\partial t}} = -{1\over 2} {{\partial^2 \vec \psi}\over{\partial x^2}} 
+\alpha ({\vec\psi}^\dagger{\vec\psi}){\vec\psi} - V_0 {\rm sn}^2 (x)\, {\vec\psi}.
\label{eom}
\ee

If $U$ is a $2\times 2$ unitary matrix,  
\be
{\vec\psi'} \equiv U {\vec\psi} 
\ee
is also a solution to the equation of motion (\ref{eom}) \cite{Park}.
This observation will allow us to construct
nonstationary solutions from the stationary solutions of Deconinck {\sl et al.}
\cite{Porter}.

{\bf III. Construction of the Nonstationary Solutions}

Nonstationary solutions can be constructed for both $\alpha = +1$ and $\alpha = -1$.
For $\alpha = -1$, any two atoms attract each other, regardless of whether or not
they belong to the same condensate.  Thus, the nonstationary solutions are likely
unstable against collapse for $\alpha = -1$.  We will therefore restrict our attention
to the case $\alpha = +1$ for the remainder of the paper.

Let $A$ and $B$ be arbitrary nonnegative real numbers.  From Ref. \cite{Deconinck_1}
we obtain three types of stationary solutions to Eq. (\ref{eom}):

\noindent Type I with

\be
\vec\psi = \left(\matrix{ A\, {\rm cn} x \cr B\, {\rm dn} x\, e^{i(1-k^2)t/2}\cr}\right ) 
\exp \left [ -i\left({1\over 2} + A^2 + B^2 \right ) t \right ]
\ee
and
\be
V = (k^2 + A^2 + k^2B^2) {\rm sn}^2 x;
\ee

\noindent Type II with

\be
\vec\psi = \left(\matrix{ A\, {\rm sn} x\cr B\, {\rm dn} x\, e^{it/2}\cr}\right ) 
\exp \left \{ -i\left[{1\over 2}(1+k^2) + A^2 + B^2 \right ] t \right\}
\ee
and
\be
V = (k^2 - A^2 + k^2 B^2) {\rm sn}^2 x + A^2;
\ee

\noindent and lastly Type III with

\be
\vec\psi = \left(\matrix{ A\, {\rm sn} x\cr kB\, {\rm cn} x\, e^{ik^2t/2}\cr}\right ) 
\exp \left \{ -i\left[{1\over 2}(1+k^2) + A^2 + k^2 B^2 \right ] t \right\}
\ee
and
\be
V = (k^2 - A^2 + k^2 B^2) {\rm sn}^2 x + A^2.
\ee

We now apply a unitary transformation to these solutions:  we set $\vec\psi' = U(\theta) \vec\psi$, where
\be\label{unitary}
U(\theta) = \left (\matrix{\cos\theta & \sin\theta\cr -\sin\theta & \cos\theta}\right ).
\ee
Formally, we can think of $\vec\psi$ as being the wave function of a spin-1/2 particle.
$U(\theta)$ then represents a rotation in spin space through an angle $\beta = -2\theta$ about the $y$-axis.
The external potential is unchanged by the transformation \cite{footnote1}.

We may confine our attention to the interval $0\le \theta < \pi$ 
because $U(\pi)\vec\psi' = - \vec\psi'$.  In fact, it is possible to reduce
this interval.  To see this, let $\vec\Psi(x, t; \theta) = \vec\psi'(x,t)$ and note
that $\vec\Psi(x,t; \theta +\pi/2) = U(\pi/2) \vec\Psi(x,t; \theta) = 
(\vec\Psi_2(x,t; \theta), -\vec\Psi_1(x,t; \theta))^T$.  Now if $(\psi_1, \psi_2)^T$
is a solution to the equation of motion (\ref{eom}), then so is 
$(e^{i\phi_1}\psi_1, e^{i\phi_2}\psi_2)^T$ for arbitrary real constants $\phi_1$ and $\phi_2$.
Moreover, these solutions have the same physical meaning.  
Apart from a switch in the labels 1 and 2 and an irrelevant phase change, therefore,
$\vec\Psi(x,t; \theta +\pi/2 )$ and $\vec\Psi(x,t;\theta)$ are identical, and it suffices
to consider the interval $0\le \theta \le \pi/2$.
There is a symmetry about $\theta = \pi/4$ that enables us to further pare down
the range of $\theta$ we must consider.  We begin with the observation that
\be
U\left ({\pi\over 2} - \theta \right ) 
 = \left (\matrix{0 & 1 \cr 1 & 0}\right ) U\left ( \theta \right ) \left (\matrix{-1 & 0 \cr 0 & 1}\right ).
\label{identity}
\ee
For Types I, II and III, $\vec\psi$ has the form $(f_1(x)e^{-i\omega_1 t}, f_2(x)e^{-i\omega_2 t})^T$.
Employing the identity (\ref{identity}), we obtain
\be
\vec\Psi\left (x,t;{\pi\over 2}-\theta\right) = \exp\left [ -i\pi\left({{\omega_2}\over{\omega_2 - \omega_1}}\right )\right ]
\left (\matrix{0 & 1 \cr 1 & 0}\right ) \vec\Psi\left (x,t - {{\pi}\over{\omega_1 - \omega_2}};\theta\right).
\ee
This means that except for a switch in the labels 1 and 2 and an irrelevant phase factor,
$\vec\Psi(x,t;{\pi\over 2}-\theta)$ and $\vec\Psi(x,t - {{\pi}\over{\omega_2 - \omega_1}};\theta)$ are the
same. Thus, we need only consider the interval $0\le \theta \le \pi/4$.

$\vec\psi' = U(\theta)\vec\psi$ is a nonstationary solution for $0 < \theta \le \pi/4$
if $\vec\psi$ is a stationary solution of Type I, II or III.  Other types of stationary
solution are constructed in Ref. \cite{Deconinck_1}: There
are solutions in which both $\psi_1$ and $\psi_2$ are proportional to 
the same Jacobian elliptic function, as well as solutions in which 
$\vert\psi_i\vert^2$ is a linear function of a Jacobian elliptic function for $i=1$ and 2.
However, when a unitary transformation is applied to solutions of these
types, the result is simply another stationary solution of the same kind.  
Therefore, we will not consider these solutions further.

{\bf IV. Physical Interpretation of the Nonstationary Solutions}

{\it A. The Trigonometric Limit $k=0$}

In the $k=0$ limit, the Jacobian elliptic functions reduce to trigonometric functions.  
For solutions of Type I, the density of condensate 1 is
\be
n_1'\equiv\vert \psi_1'\vert^2 = A^2\cos^2\theta \cos^2 x + B^2\sin^2\theta 
+ AB \sin2\theta \cos x\cos(t/2),
\ee
while the density of condensate 2 is given by
\be
n_2'\equiv\vert \psi_2'\vert^2 = A^2\sin^2\theta\cos^2 x + B^2\cos^2\theta 
- AB \sin2\theta \cos x\cos(t/2).
\ee
$n_1'$ and $n_2'$ oscillate in time with period $T=4\pi$ (in unscaled physical
units $T = 4\pi\mu/\hbar q^2$). The potential $V= A^2\sin^2 x$ is an optical
potential with minima at the points $x = m \pi$, where $m$ is any integer.  We
divide the lattice of potential minima into two sublattices: sublattice 1 with
even $m$, and sublattice 2 with odd $m$. The total condensate density
$n'\equiv n_1' + n_2' = A^2\cos^2 x + B^2$ is independent of time, and its
maxima occur at the potential minima.   At time $t=0$, the global maxima of
$n_1'$ are on sublattice 1, while at time $t=T/2$, they are on sublattice 2
(Fig.~1).  Naturally,  at time $t=T$, the global maxima of $n_1'$ have
returned to sublattice 1.   Note that the global maxima reside on one of the
two sublattices for all times $t$, and their locations change discontinuously
as time passes. The global maxima of $n_2'$ also oscillate between sublattices
1 and 2, but the oscillations of $n_2'$ lag those of $n_1'$ by half a period. 
When the condensates move between the two sublattices in this way, we will say that they undergo
sublattice oscillations.

In the $k=0$ limit, the solutions of Types I and II become identical, apart
from a  translation of both $\vec\psi'$ and $V$ through $\pi/2$.  The Type III
solution reduces to a stationary solution already studied by Deconinck {\sl et
al.} \cite{Deconinck_1} and will not be considered further here.

{\it B. Solutions with $0 < k < 1$}

We now turn to the nature of the solutions for $0<k < 1$.  Explicitly, the Type I solution is
\be
\vec\psi' = \left(\matrix{ A\, \cos\theta {\rm cn} x + B\, \sin\theta {\rm dn} x\, e^{i(1-k^2)t/2}\cr
 -A\, \sin\theta {\rm cn} x + B\, \cos\theta{\rm dn} x\, e^{i(1-k^2)t/2}\cr}\right ) 
\exp \left [ -i\left({1\over 2} + A^2 + B^2 \right ) t \right ].
\label{Type_I}
\ee
The potential minima appear on the lattice of points $x=2mK$, where $m$ is any integer.  
We again divide the lattice of potential minima
into two sublattices: sublattice 1 with even $m$, and sublattice 2 with odd $m$.
The two condensates execute sublattice oscillations with temporal
period $T=4\pi/(1-k^2)$, as shown in Fig.~2.

The nature of the solutions of Types II and III is more complex.  We will 
continue to refer to the set of points $x=2mK$ with integer $m$ as the lattice, and to divide
this lattice into sublattices 1 and 2.  The potential minima are on the 
lattice for $0 \le A < k\sqrt{1+B^2}$.  For $A>k\sqrt{1+B^2}$, on the other hand,
the potential maxima are on the lattice, and the minima lie on the set of points
$x=lK$, where $l$ is any odd integer.

Let $<n_j'>$ be 
the temporal average of $n_j'$.  For solutions of Types II and III, 
we say that the motion of condensate $j$ is of type $\alpha$ ($\beta$)
if the maxima of $<n_j'>$ are located at the odd (even) multiples of $K$.
The motion
of the two condensates will be said to be of type $\gamma\delta$ if the 
motion of condensate 1 is of type $\gamma$ and the motion of condensate 2 is of type $\delta$,
where $\gamma$ and $\delta$ can be either $\alpha$ or $\beta$.

We begin by considering the solutions of Type II.  
If the motion of condensate $j$ is of type $\beta$,
the maxima of $n_j'$ smoothly oscillate from side to side as time passes
and the oscillations of adjacent maxima in $n_j'$ are $180^\circ$ out of phase.
Let $N_l^{(j)}$ be the number of atoms of species $j$ on the interval $(l-1)K<x<(l+1)K$
for all odd integers $l$.
If the motion of condensate $j$ is of type $\alpha$, then
$N_{4m+1}^{(j)}$ oscillates periodically in time for each integer $m$.  
$N_{4m-1}^{(j)}$ also oscillates
periodically in time, but the oscillations of $N_{4m-1}^{(j)}$
lag those of $N_{4m+1}^{(j)}$ by half a period.
For motion of both types $\alpha$ and $\beta$, the period of 
the temporal oscillations $T$ is $4\pi$.

As shown in Fig.~3, motion of types $\alpha\alpha$, $\alpha\beta$, 
and $\beta\beta$ occurs for different ranges of the parameters $\theta$ and
$A/(kB)$.  (Motion of the type $\beta\alpha$ occurs in part of the region with 
$\pi/4 < \theta \le \pi/2$; these solutions are mapped to the $\alpha\beta$ sector in the
region with $0\le \theta < \pi/4$ by the symmetry transformation discussed in 
Sec. III.)  In the $\beta\beta$ sector of the
\lq\lq phase diagram," the maxima of both $n_1'$ and $n_2'$ reside in the potential
wells.  The maxima of $n_1'$ oscillate from side to side within the
wells, and the oscillations of adjacent maxima are $180^\circ$ out of phase (see Fig.~4).  The maxima of $n_2'$ oscillate in the same fashion,
but the temporal oscillations of $n_2'$ lag those of $n_1'$ by half a period.

If $A/(kB)$ is greater than
both $\cot\theta$ and $\sqrt{1+B^{-2}}$, the motion is of type $\alpha\alpha$,
and the potential minima are at the points $x=lK$, where $l$ is odd.  
The number of atoms of condensate 1 in the potential
well centered on the point $x=(4m+1)K$ (i.~e., $N_{4m+1}^{(1)}$)
oscillates periodically in time for each integer $m$.  
The number of atoms of condensate 1 in the potential
well centered on the point $x=(4m-1)K$ (i.~e., $N_{4m-1}^{(1)}$)
also oscillates periodically in time, but the oscillations of $N_{4m-1}^{(1)}$
lag those of $N_{4m+1}^{(1)}$ by half a period.  
The motion of condensate 2 is analogous to that of condensate 1, except that
it lags that of condensate 1 by half a period.  Finally, the total condensate density $n'=n_1'+n_2'$ depends on position but is independent of time.

For $k\le 1/\sqrt{2}$, the motion just described is simply a sublattice oscillation,
and the global maxima of $n_1'$ and $n_2'$ are always located at potential minima.  
A curious but interesting type of motion can occur for $k>1/\sqrt{2}$, though.
In this case, for certain values of the parameters, as a global maximum 
in $n_j'$ grows in amplitude,
it is initially located at the potential minimum.  However, as its height continues to increase,
the global maximum can split into two global maxima that are
located to either side of the potential minimum (see Fig.~5).

In the region
of the phase diagram in which $1 < A/(kB) < \sqrt{1+B^{-2}}$ and $0\le \theta\le \pi/4$, 
motion of types $\alpha\alpha$ and $\alpha\beta$ occurs, but the maxima of the
total condensate density $n'$ are located at {\it maxima} of the potential.
These solutions are therefore expected to be unstable, and this is confirmed
by our numerical simulations.

We now turn to the nature of the solutions of Type III.    
In motion of both types $\alpha$ and $\beta$, the maxima and minima
of $n_j'$ oscillate continuously from side to side, and all of maxima and minima remain
in phase with one another.  The phase diagram for the 
solutions of Type III is given by Fig.~3, just as it was for the solutions of Type II.  In the region
of the phase diagram in which $1 < A/(kB) < \sqrt{1+B^{-2}}$ and $0\le \theta\le \pi/4$,
the maxima of the total condensate density $n'$ are located at maxima of the potential,
and the solutions are expected to be unstable \cite{expectation}.
Outside this region, the maxima of $n'$ coincide with the minima of the potential.
If $A/(kB)$ is greater than both $\cot\theta$ and $\sqrt{1+B^{-2}}$, the motion
is of the type $\alpha\alpha$ and the maxima of both $n_1'$ and $n_2'$ reside
in the potential wells.  The maxima of $n_1'$ oscillate continuously
from side to side and in phase with one another (see Fig.~6).
The oscillations of the two condensates are 
$180^\circ$ out of phase, and, as a result, the total condensate density $n'$
does not depend on time.  Qualitatively speaking, the same type of intra-well oscillation
occurs in the $\beta\beta$ sector of the phase diagram.  The two condensates
oscillate in phase with one another in the $\alpha\beta$ sector of the phase diagram,
but the maxima of $<n_1'>$ coincide with the minima of $<n_2'>$.
For all solutions of Type III, the period of oscillation $T=4\pi/k^2$.

For solutions of both Types II and III, an interesting special case is obtained for $A/(kB) = \sqrt{1+B^{-2}}$.  The external potential $V$ is simply a constant in this case.  
Motion of type $\alpha\beta$ occurs if $\theta < \cot^{-1}(\sqrt{1+B^{-2}})$;
otherwise, the motion is of type $\alpha\alpha$.

{\it C. The Hyperbolic Limit $k=1$}

The Jacobian elliptic functions reduce to hyperbolic functions for $k=1$.  The Type I solution
becomes a stationary solution studied in Ref. \cite{Deconinck_1}.  The Type II and III solutions, on the other hand,
are identical for $k=1$.  Since these solutions are nonstationary, we will briefly consider
their physical interpretation.

The external potential for Type II solutions is $V=(1-A^2+B^2)\tanh^2 x + A^2$.
A method of producing a potential with a $\tanh^2 x$ spatial
dependence in an experiment is yet to be found.  The case $A=\sqrt{1+B^2}$ is therefore of particular 
interest, since $V$ is simply a constant in that case.  
For $\cot^{-1}(\sqrt{1+B^{-2}}) < \theta < \pi/4$, a bound pair of dark solitons
oscillate about the origin.  The two dark solitons move $180^\circ$ out of phase 
with one another.  For $0 < \theta < \cot^{-1}(\sqrt{1+B^{-2}})$, on the other
hand, a dark soliton and a bright soliton oscillate in phase about the
origin.  The bright soliton cannot exist in isolation since the atoms
repel one another, but the presence of the dark soliton stabilizes the bright
soliton.  Both of these kinds of solution --- oscillating dark-dark and dark-bright soliton pairs --- have previously been found by Park and Shin \cite{Park}.

This discussion suggests an alternative way of thinking about the special
cases mentioned at the close of Section IV C.  For both Types II and III,
solutions of type $\alpha\beta$ have dark-bright soliton pairs oscillating about
the lattice points, while solutions of type $\alpha\alpha$ have bound pairs of dark solitons oscillating about these points.  The dark-bright pairs oscillate in phase with each
another.  In contrast, the oscillations of the pairs of dark solitons are $180^\circ$
out of phase.  Neighboring solitons in $n_j'$ oscillate $180^\circ$ 
out of phase in Type II solutions,
whereas for Type III solutions neighboring solitons in $n_j'$ oscillate in phase with one
another.

{\bf V. Numerical Investigation of the Stability of the Nonstationary
Solutions} 

${\vec\psi'} = U(\theta) {\vec\psi}$ is a stable solution
to the equation of motion (\ref{eom}) if and only if $\vec\psi$ is a stable
solution to that equation.  This observation has two notable consequences.  
First, analytical results on the stability of stationary solutions
were obtained in Ref. \cite{Deconinck_1} and $\vec\psi$ is a stationary solution. 
Unfortunately, these analytical results are of no use here, as they
only apply if the interaction matrix $M$ is nonsingular.   
We will therefore probe the stability of the nonstationary solutions numerically.
Secondly, in our numerical work it is sufficient to investigate
the stability of our solutions for a single value of $\theta$, which we choose
to be $\pi/4$.
 
Our numerical method consists of solving the equation of motion (\ref{eom})
using a fourth-order Runge-Kutta method in time with a filtered
pseudo-spectral method in space. For each run, a small amount of white noise
was added to the initial data and the time for the onset of instability $t^*$
was determined.  For all simulations, 
\bd
t^*\equiv\min_{t>0}\left\{t:\frac{|U_n(t)-U(t)|}{U_n(t)+U(t)}=0.1\right\},
\ed
where $U(t)\equiv\max_x\left\{|\psi'_1(x,t)|,|\psi'_2(x,t)|\right\}$, 
$U_n(t)\equiv\max_x\left\{|\psi'_{1n}(x,t)|,|\psi'_{2n}(x,t)|\right\}$, and $\psi'_{1n}$ and
$\psi'_{2n}$ represent the first and second components of the numerical solutions for
$\psi'_1$ and $\psi'_2$, respectively.

The parameter space
of our nonstationary solutions is too large to permit a comprehensive study of all
possibilities. Instead, we will consider specific one-parameter families of solutions, and determine how $t^*$ varies as the parameter changes.  

{\it A. Trigonometric solutions close to the uniform solution}

The class of solutions considered first is trigonometric, and hence $k=0$. Setting
$\theta=\pi/4$, $A= \sqrt{2}\epsilon$, $B=\sqrt{2}$ and
changing the zero of energy, the solution of Type I becomes 
\be\label{eqn:plane}
\vec\psi' = 
\left(\matrix{
1+\epsilon\,e^{-it/2}\cos x 
\cr 
1-\epsilon\,e^{-it/2}\cos x 
\cr}
\right ) 
\ee
with
\be
V = -2-2\epsilon \cos^2 x.
\ee

\noindent As the parameter $\epsilon\rightarrow 0$, this solution approaches the stationary 
uniform solution, which is known to be stable \cite{Deconinck_1}. 
We investigated the stability of this class of
solutions for a range of values of $\epsilon$. In Fig.~7,
$t^*$ is plotted as a function of $\epsilon$.  The instability onset time $t^*$
appears to diverge as $\epsilon$ approaches $\epsilon^c$ from above,
where $\epsilon^c \approx 0.2$.  Two 
numerical runs are not displayed in Fig.~7: we find that 
$t^*(\epsilon=0.205)>60000$ and $t^*(\epsilon=0.2)>132000$.  These results
strongly suggest that $t^*$ is infinite for $\epsilon < \epsilon^c$, and hence that
there is a one-parameter family of stable solutions given by Eq.
(\ref{eqn:plane}) with $\epsilon<\epsilon^c\approx 0.2$.

The behavior of an unstable solution with $\epsilon=1/2$  
is illustrated in Fig.~8. For
this value of $\epsilon$, the solution is far from being uniform even at early times. 
After the onset of the instability at time $t^*\approx 450$, 
the solution loses much of its coherence,
although some is retained and new, larger-scale structures develop, as seen in
Figs. 8 (c) and (d). 

{\it B. Trigonometric solutions with increasing offset}

The second class of solutions considered is also trigonometric. With
$\theta=\pi/4$, $A=1$, $B=\sqrt{-{1\over 2} - 2\epsilon}$
and $V=2\epsilon+\frac{1}{2}-\cos^2x$, the solution of Type I becomes 

\be\label{eqn:offset}
\vec\psi' = 
\left(\matrix{
\sqrt{-\frac{1}{4}-\epsilon}+\frac{e^{-it/2}}{\sqrt{2}}\cos x 
\cr 
\sqrt{-\frac{1}{4}-\epsilon}-\frac{e^{-it/2}}{\sqrt{2}}\cos x 
\cr}
\right ) .
\ee

\noindent In general, we define the offset of a solution to be the smallest value
that $\min[n_1'(x,t), n_2'(x,t)]$ takes on for all $x$ and $t$.
The offset of the solution (\ref{eqn:offset}) increases as $\epsilon\rightarrow -\infty$. It
was shown in Ref.~\cite{Deconinck_1} that increased offset stabilizes some stationary 
solutions. This motivated us to consider
the class of solutions (\ref{eqn:offset}) with increasingly more negative
$\epsilon$. In Fig.~9, 
the instability onset time $t^*$ is shown for
a range of $\epsilon$ values.  $t^*$ remains finite for $-\epsilon$ as large
as 10, and it seems likely that $t^*$ is finite for all $-\epsilon < \infty$.
If this is indeed the case, the solutions (\ref{eqn:offset}) are all unstable. 

{\it C. A class of elliptic solutions}

The third class of solutions we consider is not trigonometric, i.~e., $k\neq 0$. In
all numerical runs, $k=0.999$, and so the solutions and the external potential are far from trigonometric.  With
$\theta=\pi/4$, $A=\epsilon$ and $B=1+\epsilon/k$, the solution of Type II is

\be\label{eqn:dn}
\vec\psi' = 
\left(\matrix{
\frac{\epsilon}{\sqrt{2}}e^{-i \omega_1 t}\sn x+
\frac{1}{\sqrt{2}}\left(1+\frac{\epsilon}{k}\right)e^{-i \omega_2 t}\dn x
\cr 
-\frac{\epsilon}{\sqrt{2}}e^{-i \omega_1 t}\sn x +
\frac{1}{\sqrt{2}}\left(1+\frac{\epsilon}{k}\right)e^{-i \omega_2 t}\dn x
\cr}
\right),
\ee

\noindent with $\omega_1=(1+k^2)/2+(1+\epsilon/k)^2$, 
$\omega_2=k^2/2+(1+\epsilon/k)^2$, and an appropriate additive shift of the
potential. As $\epsilon\rightarrow 0$, this solution approaches a stationary solution 
which has been proven to be linearly stable in the case of a nonsingular
interaction matrix \cite{Deconinck_1}.  This motivates the consideration of the
family of solutions (\ref{eqn:dn}) with parameter $\epsilon$. In
Fig.~10, the instability onset time $t^*$ is shown for different
values of $\epsilon$. The behavior of  $t^*$ is quite interesting, as it displays one
local maximum and one local minimum in the range of $\epsilon$ considered. It is
possible other local extrema exist for values of $\epsilon>1.5$, but solving
the equation of motion numerically becomes progressively more difficult as
$\epsilon$ increases because the nonlinearity grows stronger. 
As $\epsilon$ approaches $\epsilon^c\approx 0.375$ from above, $t^*$
appears to diverge. Thus, there seems to be a whole band of stable nonstationary solutions 
with the stationary solution as a limiting case.  

For values of $\epsilon>\epsilon^c$, the solution (\ref{eqn:dn}) is unstable, and
interesting structures appear after the onset of the instability. Figs.~11 - 13 
illustrate various aspects of this behavior for $k=0.999$ and $\epsilon=1/2$. From these figures, we see that the solution
becomes unstable shortly before $t=1640$, 
and then is modulated in time with a period long compared to that of the intra-well oscillations.
The amplitude of the density oscillations within the potential wells
varies by up to a factor of five.
After about three periods of the modulation, the modulated state in turn becomes unstable 
shortly after time $t=2600$. 

These phenomena are also readily observed in the Fourier transform of the solution, which is
shown in Fig.~13. 
Note the change in the spectrum during the modulated stage of the time evolution:
the amplitudes of the Fourier coefficients change, but the number of
activated modes remains approximately constant.  
The modulated state itself becomes
unstable after a few periods of
the modulation, and a band of approximately 60 unstable modes is excited, resulting
in an apparent loss of coherence in the density.

\vfill\eject

{\bf VI. Summary}

In this paper, we studied the dynamics of two-component Bose-Einstein condensates in periodic
potentials in one dimension.  Elliptic potentials which
have the sinusoidal optical potential as a special case were considered.  
We constructed exact 
nonstationary solutions to the mean-field equations of motion by performing 
a unitary transformation on previously-known stationary solutions.
Among the solutions are two types of temporally-periodic solutions ---
in one type of solution there are condensate oscillations between neighboring potential wells and in the other type
the condensates oscillate from side to side within the wells.
Our numerical studies of the stability of these
solutions suggests the existence of one-parameter families of stable
solutions for both sinusoidal optical potentials and for elliptic potentials.

R. M. B. thanks J. Roberts for useful discussions; B. D. and J. N. K.
acknowledge support from the National Science Foundation (DMS-0139093 and
DMS-0092682 respectively).

\vfill\eject

{\bf Figure Captions}

{1.} $n_1'$ (shown in dark gray) 
as a function of space and time in the optical 
potential (shown in light gray). The parameter
values for this case are $k=0$, $A=B=1/\sqrt{2}$ and $\theta=\pi/8$. Note that
the temporal oscillations of $n_2'$ lag those of $n_1'$ by half a period.

{2.} $n_1'$ for a solution of Type I as a function of 
space and time. The external
potential $V$ is shown at the rear of the figure. The parameter
values for this case are $k=0.99$, $A=B=1/\sqrt{2}$ and $\theta=\pi/8$. As
before, the temporal oscillations of $n_2'$ lag those of $n_1'$ by half a period.

{3.} A \lq\lq phase diagram" showing the types of 
motion which occur for different values of the parameters $\theta$ and
$A/kB$.  This figure applies to solutions of both Types II and III.
For $0 \le A/kB < \sqrt{1+B^{-2}}$, the potential minima are on the 
lattice of points $x=2mK$, where $m$ is an integer.  
For $A/kB>\sqrt{1+B^{-2}}$, on the other hand,
the potential maxima are on the lattice, and the minima lie on the set of points
$x=lK$, where $l$ is any odd integer.

{4.} $n_1'$ for a solution of Type II in the $\beta\beta$ sector of the phase
diagram as a function of 
space and time. The external
potential $V$ is shown at the rear of the figure.  The parameter
values for this case are $k=0.99$, $A=1$, $B=3$ and $\theta=\pi/4$. Note the
oscillations of the condensate from side to side within the potential wells.

{5.} $n_1'$ for a solution of Type II in the $\alpha\alpha$ sector of the phase
diagram as a function of 
space and time. The external
potential $V$ is shown at the rear of the figure. The parameter
values for this case are $k=0.99$, $A=2.5$, $B=1$ and $\theta=\pi/4$. Note the
periodic splitting of the peaks in $n_1'$ as the overall amount of condensate 1 in the potential wells rises and falls.

{6.} $n_1'$ for a solution of Type III in the $\alpha\alpha$ sector of the phase
diagram as a function of 
space and time.  The external
potential $V$ is shown at the rear of the figure. The parameter
values for this case are $k=0.99$, $A=3$, $B=1$ and $\theta=\pi/8$.
 
{7.} The instability onset time $t^*$ for the solutions given by
Eq. (\ref{eqn:plane}). $t^*$ appears to diverge as
$\epsilon$ approaches $\epsilon^c\approx 0.2$ from above.

{8.} The onset of the instability for the trigonometric solution
(\ref{eqn:plane}) with $\epsilon=1/2$ in a computational domain of size 
$L=8 \pi$.  Figures (a) and (b) show the
density of the first and second condensate components, respectively.  Figures (c) and
(d) are the corresponding gray scale plots.  Regions of high (low) density are shaded black
(white).

{9.} The instability onset time $t^*$ for the solutions given 
by Eq. (\ref{eqn:offset}).

{10.} The instability onset time $t^*$ for the solutions given 
by Eq. (\ref{eqn:dn}).  $t^*$ appears to diverge as $\epsilon$ approaches 
$\epsilon^c \approx 0.375$ from above.

{11.} The behavior of the elliptic solution (\ref{eqn:dn})
with $\epsilon=1/2$, $k=0.999$, $L=16K(0.999)$ and $\xi=x+L/16$,
immediately after the onset of the instability.  Figures (a) and (b) show the
densities of the first and second condensates, respectively.
Figures (c) and (d) are the corresponding gray scale plots of these quantities.

{12.} The behavior of the elliptic solution 
(\ref{eqn:dn})  with
$\epsilon=1/2$, $k=0.999$, $L=16K(0.999)$ and $\xi=x+L/16$, 
from before the onset of the instability
until well after the end of the modulated phase of the motion.  The left and right figures are
gray scale plots of the densities of the first and second condensates, respectively.

{13.} Numerical Fourier spectrum of the solution (\ref{eqn:dn}) 
with $\epsilon=1/2$ and
$k=0.999$. Only 60 of the 512 modes used are shown. The left and right
figures are the Fourier transforms of $n_1'$ and $n_2'$, respectively.


\begin{thebibliography}{99}
\bibitem{footnote1} $\vec\psi' = U(\theta) \vec\psi$ is not the most general unitary transformation,
but the most general transformation only adds phase changes to $\vec\psi_1$ and $\vec\psi_2$
and a change in the zero of time.
\bibitem{expectation} This expectation is borne out by our numerical
integrations of the equations of motion.
\end{thebibliography}
\end{document}